\documentclass[12pt]{article}
\begin{document}
\title{\bf Spin Angular Momentum Imparted by Gravitational Waves}
\author{M.Sharif\\
Department of Mathematics, University of the Punjab,\\
Quaid-e-Azam Campus, Lahore-54590, Pakistan.}

\date{}
\maketitle

\begin{abstract}
Following the demonstration that gravitational waves impart linear
momentum, it is argued that if they are polarized they should impart
angular momentum to appropriately placed 'test rods' in their path.
A general formula for this angular momentum is obtained and used to
provide expressions for the angular momentum imparted by plane and
cylindrical gravitational waves.
\end{abstract}

While the plane [1] and cylindrical [2] gravitational wave metrics
are exact solutions of Einstein's field equations (and hence have a
zero stress-energy tensor) one would expect them to carry energy. To
prove that they do so it was shown [2,3,4] that they impart linear
momentum to rest particles in their path. By the corresponding
argument [5] for the electromagnetic waves, if they are polarized
and possess a net spin, they could be expected to impart angular
momentum to 'test rods' (the 1-dimensional extension of test
particles) in their path. A formula for the linear momentum imparted
to test particles in an arbitrary spacetime had been obtained [6] by
using the extended pseudo-Newtonian ($e\psi$N)-formalism. It was
seen that the formula gave the same result as the previous analysis
for the plane gravitational wave [3] and an exact expression for he
earlier approximate results for cylindrical gravitational waves
[2,4]. Here we use the $e\psi$N-formalism to obtain a formula for
the angular momentum imparted to test rods in arbitrary spacetimes,
which is then applied to the plane and cylindrical gravitational
wave metrics.

The essence of the $\psi$N-formalism [7] was that the tidal
acceleration vector for a test rod, represented by the separation
vector $l^{\mu}$ moving along a world line with timelike tangent
vector $t^{\nu}$,
\begin{equation}
A^{\mu}=-R^{\mu}_{\nu \rho\pi}t^{\nu}l^\rho t^{\pi},
\quad(\mu,\nu,...=0,1,2,3)
\end{equation}
defines a preferred frame along which $A^{\mu}$ is maximized. In
this frame the quantity whose directional derivative along the
preferred direction is $mA^{\mu}$ (where $m$ is the rest-mass of a
test particle) gives an operationally well defined relativistic
generalization of the gravitational force, $F^{\mu}$:
\begin{equation}
l^{\nu}F^{\mu}_{;\nu}=mA^{\mu}.
\end{equation}
In the preferred frame one could replace the spcetime indices by
purely spatial indices $i,j,...=1,2,3$.

Whereas the original formalism depended on the existence of a
timelike Killing vector, the $e\psi$N-formalism [8] allowed for time
varying spacetimes. In this case $F^{\mu}$ can have a zero component
as well. The quantity whose proper time derivative is $F^{\mu}$ is
the 4-vector momentum for the test particle. The spatial components
of this vector give the momentum imparted to test particles as
defined in the preferred frame (in which $g_{0i}=0)$,
\begin{equation}
p_{i}=m\int(\ln\sqrt{g_{00}})_{,i}d\tau
\end{equation}
Thus the total energy of the test particle would be
$(m^2-p^ip^jg_{ij})^{1/2}$. As such the significance of $p_{0}$,
given by
\begin{eqnarray}
p_{0}&=&m\int[(\ln Af)_{,0}-
g^{ij}_{,0}g_{ij,0}/4A]d\tau,\nonumber\\
A&=&(\ln\sqrt{-g})_{,0},\quad g=\det(g_{ij}),\quad
f=(g_{00})^{\frac{1}{2}},
\end{eqnarray}
was not clear from the analysis.

Consider a test rod of length $\lambda$ in the path of a
gravitational wave whose preferred direction (in the above sense) is
given by $l^{i}$ in the preferred reference frame. Clearly the rod
will acquire maximum angular momentum from the wave if it lies in
the plane given by $e_{ojkl}l^{l}$, where $e_{\,\mu\nu\rho\pi}$ is
the totally skew fourth rank tensor. Thus the spin vector will be
given by [9].
\begin{equation}
S^{\mu}=\frac{1}{2}e^{\mu j k \nu}e_{jkl}l^{l}p_{\nu}.
\end{equation}
In the preferred frame the spin vector will be proportional to
$l^i$, so that
\begin{equation}
S^i=p_{0}l^i.
\end{equation}
(Here the spin can be taken to be negative if the sign of the
preferred direction is reversed.) Taking the magnitude of the spin
vector the angular momentum can be defined. Thus the maximum
angular momentum imparted to the test rod, obtained when it lies
in the plane perpendicular to the preferred direction, is
\begin{equation}
s=p_0\lambda=m\lambda\int[(\ln
Af)_{,0}-g^{ij}_{,0}g_{ij,0}/4A]d\tau.
\end{equation}
Hence the physical significance of the zero component of the
momentum 4-vector would be that it provides an expression for the
spin imparted to a test rod in an arbitrary spacetime.

The metric for plane fronted gravitational waves [4] is
\begin{eqnarray}
ds^2&=&dt^2-dx^2-L^2(t,x)[(\cos h2\beta+\sin h2\beta \cos 2\vartheta)dy^2 \nonumber\\
&+&(\cosh 2\beta-\sinh 2\beta \cos\vartheta)dz^2-2\sinh2\beta
\sin2\vartheta dydz],
\end{eqnarray}
where $L,\beta$ and $\vartheta$ are arbitrary functions of $u=t-x.$
In the case $\vartheta=0,\pi/2$ we get linearly polarized waves,
while in the case $\beta'=0\neq\vartheta'$ we get circularly
polarized waves. In general they are called elliptically polarized
[1]. The zero component of the momentum 4-vector for this line
element is $p_{0}$=constant. Hence the spin angular momentum
imparted to a test rod by it is also constant. Unfortunately the
formalism does not tell us the value of this constant. It would have
to be determined by extraneous physical considerations. Thus it
would have to be zero for linearly polarized waves and non-zero for
circularly polarized waves on physical grounds. Thus only
consistency of our analysis has so far been demonstrated. The fact
that $\dot{p}_{0}$ above is a direct consequence of the vacuum
Einstein equations. It is to be noted that even with the vacuum
Einstein equations holding, it is not necessary that
$p_{0}=constant$. This is seen in the case of cylindrical
gravitational waves which we discuss next.

The metric for cylindrical gravitational waves is
\begin{equation}
ds^2=e^{2(\gamma-\psi)}dt^2-e^{2(\gamma-\psi)}d\rho^2
-\rho^2e^{-2\psi}d\phi^2-e^{2\psi}dz^2,
\end{equation}
where $\gamma$ and $\psi$ are arbitrary functions of the time and
radial coordinates, $t$ and $\rho$ and are given by
\begin{eqnarray}
\psi&=&AJ_{0}(\omega \rho)\cos \omega t+BY_{0}(\omega \rho)\sin
\omega t, \nonumber\\
\gamma&=& \frac{1}{2}\omega \rho[(A^2J_0J'_0-B^2Y_0Y'_0)\cos 2
\omega t- AB(J_0Y'_0+Y_0J'_0)\sin 2\omega t \nonumber\\
&-&2(J_0Y'_0-Y_0J'_0) \omega t],
\end{eqnarray}
where $A$ and $B$ are arbitrary constants corresponding to the
strength of the cylindrical gravitational waves, $J_{0}$ and
$Y_{0}$ are the Bessel function and the Neumann function of zero
orders respectively. Here prime denotes differentiation with
respect to $\omega\rho,\omega$ being the angular frequency. To
avoid singularities at the source, following Weber and Wheeler
[2,4], we take $B=0$.

The zero component of the momentum 4-vector is
\begin{equation}
p_{0}=-m[1+AJ_{0}/\omega \rho J'_0)\ln|1-2\omega\rho AJ'_0 \cos
\omega t|]
\end{equation}
Hence the spin angular momentum imparted to a test rod by it is
\begin{equation}
s=-m \lambda[(1+AJ_{0}/\omega \rho J'_{0})\ln|1-2 \omega \rho
AJ'_{0}\cos \omega t|]+constant.
\end{equation}
It is clear that for a zero amplitude wave $(A=0)$ there would be no
angular momentum imparted to the test rod, as one would naturally
require. Hence the constant of integration must be zero here. Notice
that we must require that $A$ be such that $|2\omega \rho J'_0A|<1$
so that $p_0$ and $s$ remain non-singular.

There being only two exact gravitational wave solutions available we
are limited in exploring the consequences of our analysis. The plane
wave case highlighted the shortcoming of our analysis, in that it
can not determine the constant of integration without reference to
extraneous physical considerations. On the other hand, the
cylindrical wave analysis demonstrates that physical consequences
\emph{can} be deduced using the $e\psi$N-formalism.

If there is a linear superposition of spins in the gravitational
wave we would have to take the corresponding linear superposition
of the resulting expressions for the angular momentum imparted by
the waves to test rods. Thus, for a random superposition the net
angular momentum transferred would be zero. However, if the waves
have only one polarization then they must carry a net spin along
the preferred direction (positive helicity) or opposite to it
(negative helicity). In this case the spin would become apparent
through the angular momentum it imparts to the test rod.

\vspace{.5cm}

\textbf{Acknowledgement}

\vspace{.5cm}

I would like to thank Prof. Asghar Qadir for his useful discussions
and comments during the write up of this paper.
\newpage

\textbf{Reference}

\begin{description}
\item{[1]}  Misner, C.W., Thorne, K.S. and Wheeler, J.A:
\textit{Gravitation}, (W.H. Freeman, San Francisco, 1973);\\
Kramer, D., Stephani, H., Herlt, E., MacCallum, M.: \textit{Exact
Solutions of Einsterin's Field Equation}, (Cambridge University
Press, 1979).

\item{[2]} Weber, J. and Wheeler, J.A: Rev. Mod. Phys. \textbf{29}(1957)509.

\item{[3]} Ehlers, J. and Kundt, W.: \textit{Gravitation:
An Introduction to Current Research} ed. Witten, L. (Wiley, New
York, 1962).

\item{[4]} Weber, J.: \textit{General Relativity and
Gravitational Waves}, (Interscience Publishers, Inc., New York,
1961).

\item{[5]} Purcell, Edward, M.: \textit{Electricity and Magnetism:
Berkeley Course-Vol. 2}, (McGraw-Hill Book Company, 1965).

\item{[6]} Qadir, Asghar and Sharif, M.: Phys. Lett. \textbf{A167}(1992)331.

\item{[7]} Qadir, A. and Quamar, J.: \emph{Proceedings of the Third
Marcel Grossmann Meeting on General Relativity}, ed. Hu Ning (North
Holland, Amsterdam, 1983)p189;\\
Quamar,J.: Ph.D. Thesis, Quaid-i-Azam University (1984).

\item{[8]} Qadir, A. and Sharif, M.: \emph{Proceedings of the Fourth Regional
Conference on Mathematical Physics} eds. F. Ardalan, H. Arfaei and
S. Rouhani (Sharif University of Technology Press, 1993)p182; Nuovo
Cimento \textbf{B107}(1991)1071;\\
Sharif, M.: Ph.D. Thesis Quaid-i-Azam University (1991).

\item{[9]} Penrose, R. and Rindler, W.: \emph{Spinors and
Spacetime, Vol.1} (Cambridge University Press, 1984).

\end{description}

\end{document}